# Reducing semantic complexity in distributed Digital Libraries: treatment of term vagueness and document re-ranking

Philipp Mayr, Peter Mutschke, Vivien Petras
GESIS-IZ Social Science Information Centre, Bonn (Germany)
http://www.gesis.org/IZ/
E-mail: philipp.mayr|peter.mutschke|vivien.petras@gesis.org

**Abstract**

**Purpose** - The general science portal vascoda merges structured, high-quality information collections from more than 40 providers on the basis of search engine technology (FAST) and a concept which treats semantic heterogeneity between different controlled vocabularies. First experiences with the portal show some weaknesses of this approach which come out in most metadata-driven Digital Libraries (DL) or subject specific portals. The purpose of the paper is to propose models to reduce the semantic complexity in heterogeneous DLs. The aim is to introduce value-added services (treatment of term vagueness and document re-ranking) that gain a certain quality in DLs if they are combined with heterogeneity components established in the project "Competence Center Modeling and Treatment of Semantic Heterogeneity".

**Design/methodology/approach** - First, semantic heterogeneity components translate automatically between different indexing languages. This approach will have an impact on search in a scenario when the searcher uses controlled vocabularies which are cross-linked with cross-concordances. However, users usually formulate query terms freely without any vocabulary support. Empirical observations show that freely formulated user terms and terms from controlled vocabularies are often not the same or match just by coincidence. Therefore, a value-added service will be developed which rephrases the natural language searcher terms into suggestions from the controlled vocabulary, the Search Term Recommender (STR). Second, the result sets of transformed or expanded queries in distributed collections are often very large and tests show that the conventional web-based ranking methods are not appropriate for presenting heterogeneous metadata records as suitable result sets to the user. Therefore, two methods, which are derived from scientometrics and network analysis, will be implemented with the objective to re-rank result sets by the following structural properties: the ranking of the results by core journals (so-called Bradfordizing) and ranking by centrality of authors in co-authorship networks.

**Findings** - The methods, which will be implemented, focus on the query and on the result side of a search and are designed to positively influence each other. Conceptually they will improve the search quality and guarantee that the most relevant documents in result sets will be ranked higher.

**Originality/value** - The central impact of the paper focuses on the integration of three structural value-adding methods which aim at reducing the semantic complexity represented in distributed DLs at several stages in the information retrieval process: query construction, search and ranking, and re-ranking.

**Paper type** - Research paper

**Keywords:** Digital Library, Semantic Heterogeneity, Search Term Recommender, Re-Ranking, Bradfordizing, Co-Author Networks, Network Analysis

# Introduction

In the area of scientific and academic information systems, a whole array of bibliographic databases, disciplinary internet portals, institutional repositories or archival and other media type collections are increasingly accumulated and embedded in all-encompassing information systems in order to meet user requirements that demand one-stop "information fulfillment". Examples are Elsevier's Scirus portal [1], the OCLC Worldcat union catalog [2] or Tuft University's Perseus project [3].

In Germany, an ambitious project for one-stop academic search is the vascoda portal [4], a joint project between the BMBF (Federal Ministry for Education and Research) and the DFG (German Research Foundation). Vascoda provides a federated search interface for a multitude of disciplinary and interdisciplinary databases (e.g. full-text article databases, indexing and abstracting services, library catalogs) and internet resource collections.

The vascoda portal contains many information collections that are meticulously developed and structured. They have sophisticated subject metadata schemes (subject headings, thesauri or classifications) to describe and organize the content of the documents on an individual collection level. The general search interface, however, only provides a free-text search over all metadata fields without regarding the precise subject access tools that were originally intended for these information collections.

If large-scale contemporary information organization efforts like the Semantic Web [5] (see also Krause, 2006, 2007, Krause to appear) strive to provide more structure and semantic resolution of information content, how is it possible that advanced interfaces for digital libraries (DL) scale back on exactly the same issue?

Search, both in full-text collections like the Internet or more heavily structured and less diverse collections like institutional repositories, indexing databases or library catalogs as described above, only works as well as the matching between the language in queries and the language in the searched documents. If the words in the query are different from the words in a relevant document, this document will not be found. The problem of matching query terms to document terms is a result of the ambiguity or vagueness of language (Blair, 1990, 2003).

Because of the sheer size and variation of large full-text databases, this problem is not as noticeable because any query (even if they contain spelling mistakes or nonsense statements) will find documents. The problem is aggravated in collections of more restricted volume or text, i.e. repositories that contain only formal metadata, some subject description and just a link to the full-text. The issue becomes even more critical when several collections with different metadata schemes are searched at the same time – which is the case in the distributed search scenario. In this scenario, not only the matching between query and document terms is affected by language ambiguity, but also the matching between different subject-describing metadata schemes. In figure 1, we speak of vagueness 1 and vagueness 2/3 (V1 and V2/3) to denote the different areas where language ambiguity can occur. For a successful retrieval in any digital library, both levels of vagueness have to be addressed (compare Hellweg et al, 2001).

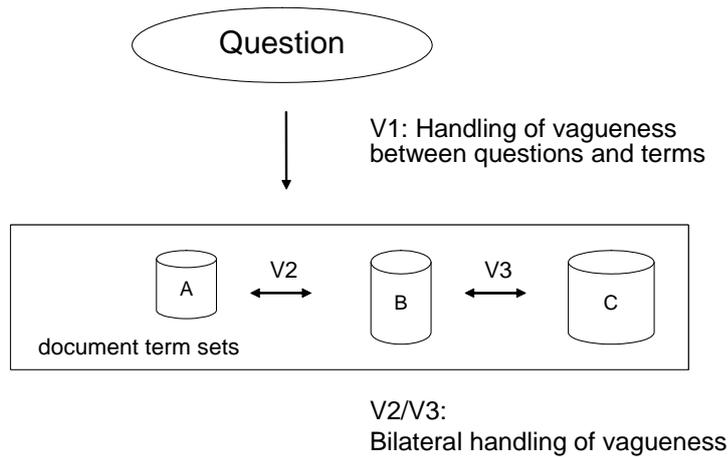

Figure 1: Two step methodology of vagueness treatment

Furthermore, the result sets of transformed or expanded queries in distributed collections are often very large and tests show that the conventional web-based ranking methods are not appropriate for the heterogeneous metadata records. Therefore, two methods, which are derived from scientometrics and network analysis, will be implemented with the objective to re-rank result sets: a) the ranking of the results by core journals (so-called Bradfordizing) and b) ranking by centrality of authors in co-authorship networks.

This paper is a description of an attempt to harness the semantic knowledge in controlled vocabularies for several stages in the information retrieval process: query construction, search and ranking and re-ranking. We briefly describe the GESIS project "Competence Center Modeling and Treatment of Semantic Heterogeneity" with the goal of creating a semantic network of terms in different controlled vocabularies (terminology mapping) in order to facilitate a seamless search across different subject-describing knowledge organization systems (KOS). At the conclusion of this project, we are now devising modules to leverage these mappings for an improved search experience for the user.

## Results from a major terminology mapping effort

Semantic integration seeks to connect different information systems through their subject metadata frameworks - insuring that distributed search over several information systems can still use the advanced subject access tools provided with the individual databases. Through the mapping of different subject terminologies, a "semantic agreement" for the overall collection to be searched on is achieved. Terminology mapping – the mapping of words and phrases of one controlled vocabulary to the words and phrases of another – creates a semantic network between the information systems carrying the advantages of controlled subject metadata schemes into the distributed digital library world.

In 2004, the German Federal Ministry for Education and Research funded a major terminology mapping initiative at the GESIS Social Science Information Centre in Bonn (GESIS-IZ) "Competence Center Modeling and Treatment of Semantic Heterogeneity" [6], which concluded this year (see Mayr/Walter, 2007a, 2007b). The task of this terminology mapping initiative was to organize, create and manage 'cross-concordances' between major controlled vocabularies (thesauri, classification systems, subject heading lists) centred around the social sciences but quickly extending to other subject areas (e.g. political science, economics, medicine or subject-specific parts of universal vocabularies). Cross-concordances are intellectually (manually) created crosswalks that determine equivalence, hierarchy, and association relations between terms from two controlled vocabularies. Most vocabularies in the project have been related bilaterally, that is, there is a cross-concordance relating terms from vocabulary A to vocabulary B as well as a cross-concordance relating terms from vocabulary B to vocabulary A (bilateral relations are not necessarily symmetrical). Other definitions and examples of crosswalks between controlled vocabularies exist in an international context (see overview in Zeng/Chan, 2004; Vizine-Goetz et al., 2004; Liang/Sini, 2006).

In November 2007, 25 controlled vocabularies from 11 disciplines were connected with vocabulary sizes ranging from 1,000 – 17,000 terms per vocabulary. To date, more than 513,000 relations in 64 crosswalks have been generated. An overview of the preliminary project results presented at the NKOS/ ECDL workshop 2007 can be found in [7].

A database including all mapped controlled terms and cross-concordance relations was built and a 'heterogeneity service' developed, a web service, which makes the cross-concordances available for other applications (see figure 2). Many cross-concordances are already implemented and utilized for the German Social Science Information Portal sowiport [8], which searches bibliographical and other information resources (including 13 databases with 10 different vocabularies and about 2.5 million references).

"take in Figure 2"

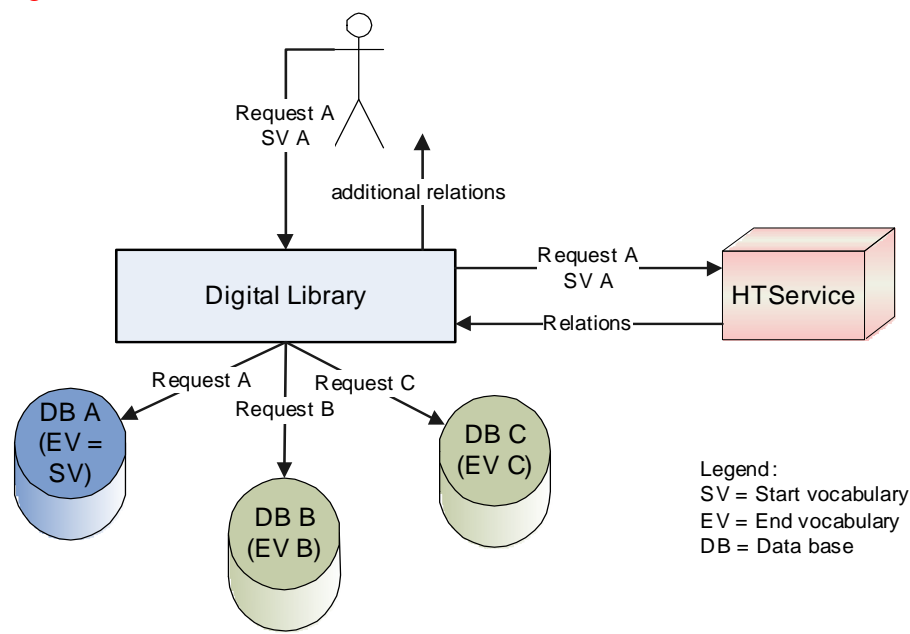

Figure 2: Heterogeneity Service (HTS)

Semantic mappings could support distributed search in several ways. First and foremost, they should enable seamless search in databases with different subject metadata systems. Additionally, they can serve as tools for vocabulary expansion in general since they present a vocabulary network of equivalent, broader, narrower and related term relationships. Thirdly, this vocabulary network of semantic mappings can also be used for query expansion and reformulation.

The following chapter introduces the concept of a search term recommender. This tool is an aid for query reformulation and reconstruction that has been adapted from human search intermediaries (e.g. reference librarians) to the automatic version for a web-based information portal.

## Search Term Recommender (STR)

Semantic mappings can reduce the problem of language ambiguity at the vagueness 2/3 layer described in figure 1 (between different information systems). However, the vagueness at the user-information system interface still hasn't been addressed.

To reduce language ambiguity at the vagueness 1 layer between query terms and document terms, another instrument is necessary to map terms at this interface. The goal of this mapping would be to "translate" the query terms of a user to the document terms of the database (or vice versa) in order to produce a match at search time. Since we are mostly concerned with information systems that contain sparse text enriched with subject describing controlled vocabularies, we propose a search term recommender system, which will propose terms from the controlled vocabularies given a specified query.

The basic parameters of a search term suggestion system are the controlled vocabulary terms that are used for document representation and the natural language keywords that are input by the searcher. The advantage of suggesting controlled vocabulary terms as search terms is that these terms have been systematically assigned to the documents so that there is a high probability of relevant and precise retrieval results if these terms are used instead of whatever natural language keywords the searcher happens to think of.

A second advantage in suggesting controlled vocabulary terms is their application in the semantic network of the cross-concordances. That is, if controlled vocabulary terms are used in search, the cross-concordances, which map these terms between different databases, can be successfully applied for distributed retrieval.

This kind of vocabulary help will hopefully also improve the search experience for the user in general. Suggesting terms on the one hand reduces the searcher's need to think of other search terms that might describe his information need. It effectively eases the cognitive load on the searcher since it is much easier for a person to pick appropriate search terms from a list than to come up with search terms by himself. On the other hand, it also helps alleviating "anchoring bias" (Blair, 2002), which is an effect that makes it harder to substantially deviate from one's original thought-of search terms and to consider different search terms or strategies.

Another consequence of term suggestion is the presentation of new or different technical expressions for a concept. This again could lead to changes in search strategy or topic, which might help in reaching the search goal. Term suggestions from several fields of research and / or information resources could also provide an overview over different areas of discussion, which deal with particular concepts (assuming different meanings or directions of thought). The result would be a different perspective on certain concepts, an effect, which can also be achieved by displaying the semantic mappings of the cross-concordances themselves.

A search term recommender is created by building a dictionary of associations between two vocabularies: (1) natural language terms and phrases from the subject area's documents in the information collection (e.g. titles, abstracts, authors) and (2) the controlled vocabulary (thesaurus terms, subject headings, classification numbers etc.) used for document representation.

In one implementation, a likelihood ratio statistic is used to measure the association between the natural language terms from the collection and the controlled vocabulary terms to predict which of the controlled vocabulary terms best mirror the topic represented by the searcher's search terms (Plaunt/Norgard, 1998; Gey et al., 1999). However, other methods of associating natural language terms and controlled vocabulary terms are possible (Larson, 1991, 1992).

In an information system with several information resources (i.e. databases) and several controlled vocabularies, a search term recommendation tool has to determine which terms from which vocabularies to suggest and how to tie the term suggestions for query construction into the semantic network of cross-concordances. Several approaches seem possible: a pivot controlled vocabulary, from which terms are suggested and mappings approached; a general suggestion pattern, which clusters similar concepts from several vocabularies; or a domain-specific approach, whereby terms and vocabularies are chosen according to the subject of interest for the searcher.

The result sets of transformed or expanded queries in distributed collections are often very large and tests show that the conventional web-based ranking methods are not appropriate for presenting heterogeneous metadata records as suitable result sets to the user. In the following chapter we propose re-ranking methods (implemented as post-search modules), which are based on structures and regularities in scientometrics and network analysis.

## Re-ranking

Compared to traditional text-oriented sorting mechanisms, our scientometric and network analysis re-ranking methods offer a completely new view on results sets which have not been implemented in heterogeneous and larger database scenarios to date. The usage of these modules should be an alternative ranking opportunity with the objective to enhance and improve the search process in general. In addition, we expect an improvement in document relevance for the top-listed documents.

### Bradford Law of Scattering and Bradfordizing

Bradford Law of Scattering and Bradfordizing have their roots in scientometrics and are often applied in bibliometric analyses of databases and collections as a tool for systematic collection management in library and information science. Fundamentally, Bradford Law says that a literature on any scientific field or subject-specific topic scatters in a typical way. A core or nucleus with the highest concentration of papers (few core journals) on a topic is followed by zones with loose concentration of paper frequency, which is described by Bradford:

> *"… if scientific journals are arranged in order of decreasing productivity of articles on a given subject, they may be divided into a nucleus of periodicals more particularly devoted to the subject and several groups or zones containing the same number of articles as the nucleus, when the numbers of periodicals in the nucleus and succeeding zones will be as $1:n:n^2$ …"* (Bradford, 1948)

Bradford Law as a general law in informetrics can be applied to all scientific disciplines and especially in a multi-database scenario in combination with semantic treatment of heterogeneity as described before. Bradfordizing (White, 1981) is an information science application of Bradford Law of Scattering which sorts/re-ranks a result set according to the identified core journals for a query. The journals for a search are ranked by the frequency of their listing in the result set (number of articles for a journal title). If a search result is bradfordized, articles of core journals are ranked ahead of the journals which contain an average number or only few articles on a topic. This method is interesting in the context of our re-ranking task because it is a robust way of sorting the central publication sources for any query to the top positions of a result set. Bradfordizing has the following values-added:
1. An alternative view on results sets which are ordered by core journals (the user is provided with documents of core journals first),
2. An alternative view on publication sources in an information space which are intuitively closer at the research process than statistical methods (e.g. best match) or traditional methods (e.g. exact match),
3. Possibly a higher topical relevance of re-ranked documents.

Additionally, re-ranking via bradfordized lists offer an opportunity to switch between term-based search and the alternative search mode browsing. Bates (2002) brings together Bradford Law and information seeking behavior.

> *"… the key point is that the distribution tells us that information is neither randomly scattered, nor handily concentrated in a single location. Instead, information scatters in a characteristic pattern, a pattern that should have obvious implications for how that information can most successfully and efficiently be sought." (Bates, 2002)*

Bates applies conceptually different search techniques (directed searching, browsing and linking) to the Bradford zones. Bates postulates the Bradford nucleus for browsing, the second zone for directed searching with search terms and further zones for linking. We focus on an automatic change from directed searching (enhanced by treatment of semantic heterogeneity) into browsing. Starting with a subject specific descriptor search we will connect the query with our heterogeneity service to transfer descriptor terms into a multi-database scenario. In the second step, the result lists from the different databases will be combined and sorted according to Bradford's method (most productive journals for a topic first). After this step we have a bradfordized list of journal articles. The next step is the extraction of a result set of all documents in the Bradford nucleus which can be delivered for browsing. This automatically generated browsing modus can be compared to Bates search technique "journal run".

The focusing on re-ranking basing on Bradfordizing is interesting because of the universal properties of the law which can be applied in a one-database scenario (e.g. Mayr/Umstätter, 2007) and a multi-database scenario like vascoda or sowiport. On a very abstract level, Bradford re-ranking can be used as a compensation method for enlarged search spaces.

However, in our application model information on the core journals is used for document ranking.

**Co-Author networks**

It is a known fact that standard search services do not meet the wealth of information material supplied by DLs. Traditional retrieval systems are strictly document oriented such that a user is not able to exploit the full complexity of information stored. Bibliographic data, for instance, offer a rich information structure that is usually "hidden" in traditional information systems. A typical example of this issue are link structures among authors, given for instance by co-author relationships, and - more importantly - the strategic position of authors within a given collaboration structure. On the other hand, the continuous growth of accessible information implies that the user is more concerned with potentially irrelevant information. Moreover, relevant information is more and more distributed over several heterogeneous information sources and services (bibliographic reference services, citation indices, full-text services).

DLs are therefore only meaningfully usable if they provide high-level search services that fully exhaust the information structures stored and, and the same time, reduce the complexity of information to highly relevant items. Due to the notion of a Semantic Web, particularly the **F**riend **of** a **F**riend (FOAF) approach, this strongly suggests the development of techniques that overcome the strict document orientation of standard indexing and retrieval methods by providing a deeper analysis of link structures and the centrality of entities in a given network structure.

This approach focuses on network analysis concepts for extracting central actors in co-author networks and ranking documents by author centrality (Mutschke, 2003). The expressiveness of co-author networks has been demonstrated in a number of scientometric studies (see e.g. Beaver, 2004). The basic approach of our model is to reason about the network structure in order to evaluate relevant authors for a particular domain and to use information on the centrality of authors within their scientific community to rank documents.

According to graph theory a co-author network in our model is described as a graph $G=(V,E)$, where $V$ is the set of vertices (authors), and $E$ the set of edges (co-authorships). A co-author network is propagated on the basis of all co-authorships that appear in a given document set, for instance the result set of a query. On social networks a number of calculations can be performed. An important structural attribute of the vertices is their *centrality*. Centrality measures the contribution of a network position to the vertices' prominence, influence or importance within a social structure. In our model we use the *Betweenness* measure. Betweenness focuses on the ratio of shortest paths a vertex lies on. An author with a high betweenness is therefore a vertex that connects many authors in the network. Betweenness is therefore seen as a measure that indicates an actor's degree of control or influence of interaction processes that construct a network structure.

Accordingly, an index of centrality within a scientific collaboration and communication structure might indicate the degree of relevance of an author for the domain in question because of his/her key position in the network. In our application model information on the centrality of authors is used for document ranking. This is done by weighting the documents retrieved by the centrality values of their authors such that the user is provided with documents of central authors.

Figure 3 visualizes the planned application of value-added services in the stages of the search process and the combination of the single components. See search term recommender (STR) in the beginning of a search and re-ranking of combined search result sets at the end of a search loop (see stage 2 and 7 in figure 3).

"take in Figure 3"

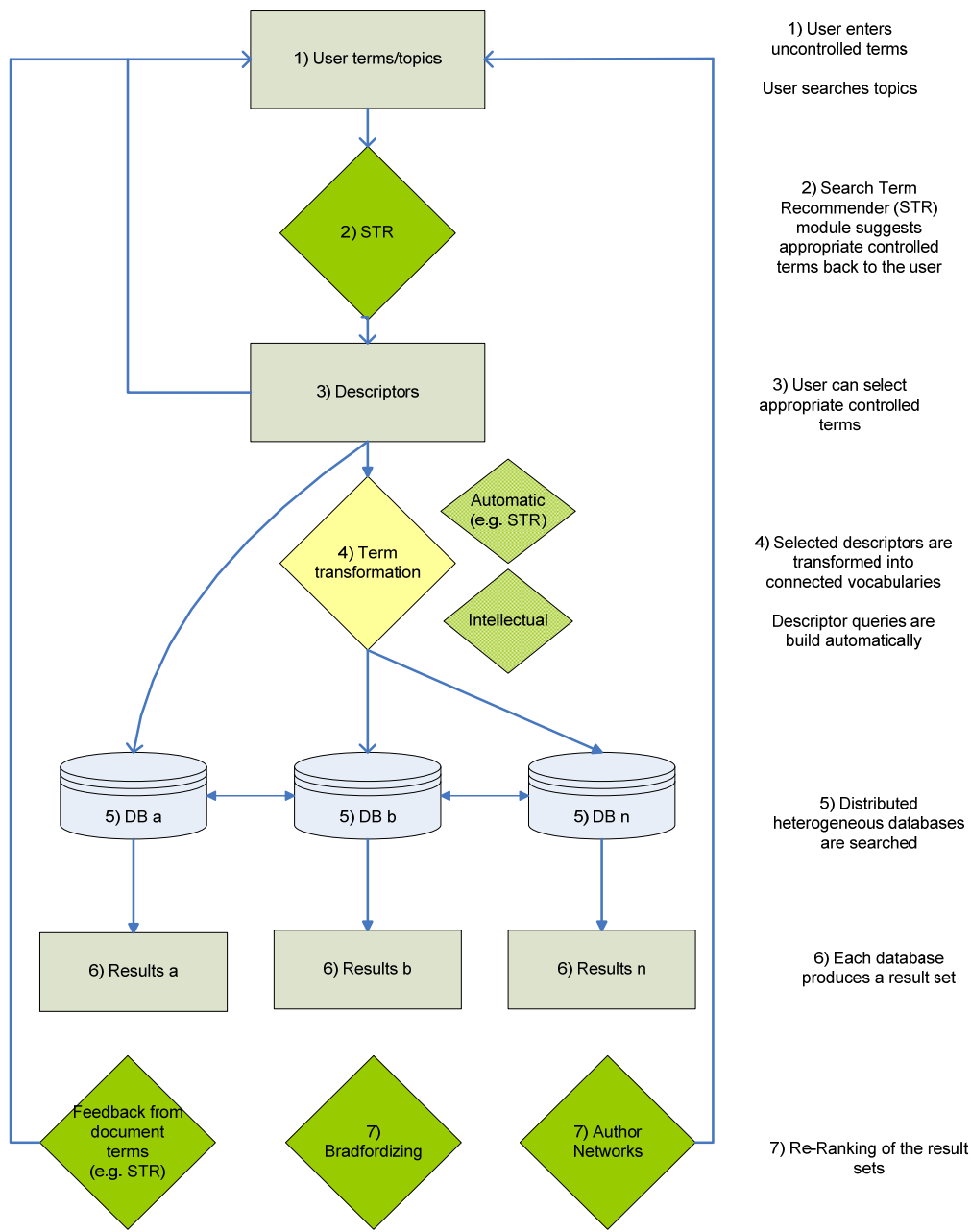

Figure 3: Combination and embedding of the modules: STR and re-ranking

## Integration

Beyond an isolated use, a combination of the approaches is promising to yield much higher innovation potential. In our model, the following scenarios are supported (e.g. combining Bradfordizing with Author Centrality as in figure 4).

The user is provided with publications which are associated with both central authors as well as core journals. From a technical point of view, the following variants are suitable which may yield different results:

- Bradfordizing as a filter for the network analysis process: central authors are evaluated within the set of documents which are associated with core journals, i.e. the result set is reduced to the core journal set before author centrality analysis is performed.
- Author centrality as a filter for Bradfordizing (the "inverse" version of the variant above): Bradfordizing is performed on the set of result set document which are assigned to central authors, i.e. the result set is reduced to "central" documents before core journals are evaluated[1].
- The "intersection" variant: core journals and central authors are first evaluated independently from one another on the basis of the whole result set. Publications that satisfy both relevance criteria (they appear in a core journal *and* their authors are central) are determined in a second step (see figure 4).

"take in Figure 4"

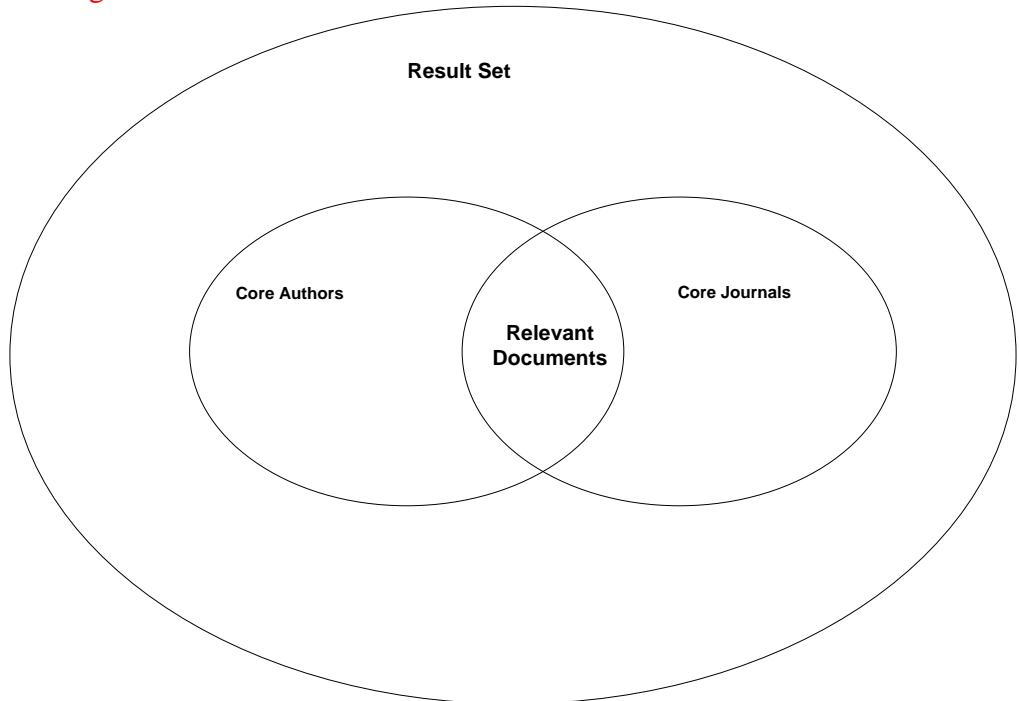

Figure 4: Intersection of core journal and central author documents

Those combination models could not only be applied to result set re-ranking but also, and in the same way, to the search term recommendation process, i.e. the usage of Bradfordizing and author centrality analysis as a filter on the collection used for the STR analysis.

Future research work should address the use of information items others than authors and journals, such as institutions, themes or citations as further value-adding functions in re-ranking methods. An important further research issue is to apply and evaluate the proposed ranking methods at the search stage in order to improve the precision of the initial result set.

## Outlook

The central impact of the paper focuses on the integration of three structural value-adding methods which aim at reducing the semantic complexity represented in distributed DLs at several stages in the information retrieval process: query construction, search and ranking and re-ranking. The integration of the models will be done using Semantic Web technologies which should enhance further insights into the usage of those techniques. The intersection of the Semantic Web world with the Digital Library world as mentioned in Krause (to appear) will hopefully result in more sophisticated analytic tools and interfaces for the presentation of information adapted to a user's needs.

---
[1] For this variant a (configurable) threshold for centrality is needed.

**Links**

1. Elsevier. "Scirus - for scientific information only." Retrieved October 2007, from http://www.scirus.com/.

2. Online Computer Library Center (OCLC). "WorldCat." Retrieved October 2007, from http://www.oclc.org/worldcat/.

3. Department of the Classics, Tufts University. "The Perseus Digital Library." Retrieved October 2007, from http://perseus.mpiwg-berlin.mpg.de/.

4. "vascoda - Entdecke Information." Retrieved October 2007, from http://www.vascoda.de/.

5. World Wide Web Consortium (W3C). (2001). "Semantic Web Activity." Retrieved October 2007, from http://www.w3.org/2001/sw/.

6. "Competence Center Modeling and Treatment of Semantic Heterogeneity." Retrieved October 2007, from http://www.gesis.org/en/research/information_technology/komohe.htm

7. Results from a German terminology mapping effort: intra- and interdisciplinary cross-concordances between controlled vocabularies. Presented at the NKOS/ECDL Workshop in Budapest Hungary. Retrieved October 2007, from http://dlist.sir.arizona.edu/2054/

8. Sowiport. Retrieved October 2007, from http://www.sowiport.de